# Is Low-Frequency-Peaked BL Lac Object OJ 287 a TeV Emitter?*

CHEN Liang[1,2,3]**, BAI Jin-Ming[1,2]

[1] *National Astronomical Observatories/Yunnan Observatory, Chinese Academy of Sciences, Kunming 650011, China*
[2] *Key Laboratory for the Structure and Evolution of Celestial Objects, Chinese Academy of Sciences, Kunming 650011, China*
[3] *The Graduate School of Chinese Academy of Sciences, Beijing 100049, China*

*Supported by the West PhD Project of the Training Program for the Talents of West Light Foundation of the CAS, and the National Natural Science Foundation of China under Grant Nos 10903025, 10778702, 10973034 and 11078008, and the 973 Program under Grant No 2009CB824800.

**Email: chenliangew@hotmail.com; baijinming@ynao.ac.cn

(Received 24 May 2010)

It is well known that there are only two low-frequency-peaked BL Lac objects (LBLs: BL Lacertae and S5 0716+714) and one flat spectrum radio quasar (FSRQ: 3C 279) among more than 30 active galactic nuclei (AGNs) with detected TeV emissions. We study the spectral energy distribution (SED) of a famous LBL OJ 287, whose light curve has a 12-y period. Using a homogeneous one-zone synchrotron + synchrotron-self Compton model, we model the quasi-simultaneous broad-band SED of OJ 287. With some reasonable assumptions, we extrapolate the model to the high state of OJ 287 and predict its $\gamma$-ray emissions. Taking into account the absorption of $\gamma$-ray by the extragalactic background light (EBL), we find that the TeV emission of OJ 287 in high state is slightly higher than the sensitivity of H.E.S.S. The study on SEDs of OJ 287 has implications to unveil the origin of jet activity during its 12-y period and the properties of EBL.

*PACS*: 95.30.Gv, 95.85.Pw, 98.62.Mw, 98.62.Nx, 98.54.Cm

Blazars are the most extreme active galactic nuclei (AGNs), whose continuum emissions are dominated by the nonthermal emissions from the relativistic jet co-aligned with our line of sight.[1-4] The spectral energy distributions (SEDs) of blazars consist of two broad bumps. The lower frequency bump is due to the synchrotron emissions of relativistic electrons in the jet. The higher frequency bump is thought to be due to the inverse Compton (IC) emissions of the same electrons (however see the hadronic model[5-8]). The soft seed photons for IC scattering may come from the synchrotron emissions (synchrotron-self Compton, SSC[2-4,9,10]) or come from the external regions (external Compton, EC[11-14]). The former (SSC) may be responsible for γ-rays in BL Lac objects (BL Lacs), while the latter (EC) would work in flat spectra radio quasars (FSRQs).

According to different peaks of the synchrotron components, BL Lacs could be classed as low-frequency peaked BL Lacs (LBL) and high-frequency peaked BL Lacs (HBL).[15,16] The synchrotron peak frequencies of FSRQs are similar to that of LBLs or even lower.[17] Generally, LBLs have the IC peaks below the GeV band and the HBLs have the IC peaks up to subTeV band. Among more than 30 AGNs detected with TeV emissions (Update sees, http://tevcat.uchicago.edu/ and/or http://www.mppmu.mpg.de/~rwagner/sources/) there are only two LBLs: BL Lacertae[18] and S5 0716+714[19], and one FSRQ: 3C 279.[20] The nature why these three low-frequency peaked blazars can emit so high energy photons, what are the physical differences between these three blazars and other low-frequency peaked blazars and other TeV HBLs are not clear.[21] It obviously needs to expand the sample of such particular blazars to unveil their nature.

OJ 287 ($z$=0.306[22]) is an LBL which has been well studied through radio to x-ray bands (see Refs.[23-26] for radio studies, Refs.[27-33] for optical, Refs.[34-37] for x-rays; and references therein). The most outstanding characteristic of OJ 287 is its 12-y period, which is discovered in optical band[27,29] and has also been confirmed in the x-ray band.[37] OJ 287 has been detected with GeV emissions by EGRET[38] and Fermi/LAT.[39,40] It has also been prospected to be a TeV source, but still not be detected.[41-43] The SED modelings were based on non-simultaneous SEDs in the previous studies (the modeled SED in Ref. [42] is quasi-simultaneous but lacks the data below the optical band). In this Letter, we study the emission properties of OJ 287 by modeling its quasi-simultaneous SED from radio to γ-ray. Furthermore, we extrapolate the model to predict its TeV emission properties in high state. Throughout the study, a cosmology with $H_0$ = 70 km·s$^{-1}$·Mpc$^{-1}$, $\Omega_m$ = 0.3 and $\Omega_\Lambda$ = 0.7 is adopted.

The red bow tie in Fig. 1 is the average spectrum from the first three months detection by Fermi/LAT.[39,42] The red stars are the quasi-simultaneous SED within these three months operation of Fermi/LAT.[44] The blue squares and bow tie are taken from previous observations[41,42]. In modeling the emissions of OJ 287, we use a simple one zone synchrotron + SSC model.[45,46] The emission region is assumed to be a homogeneous sphere with radius $R$ embedded in the magnetic field $B$. We assume the electron energy distribution (volume density) to be a break power law,

$$N(\gamma) = \begin{cases} N_0 \gamma^{-p_1}, & \gamma \leq \gamma_p, \\ N_0 \gamma_p^{p_2-p_1} \gamma^{-p_2}, & \gamma > \gamma_p. \end{cases} \quad (1)$$

The parameters in the model include $R$, $B$, electron peak energy $\gamma_p$, normalized density $N_0$, indexes $p_{1,2}$ and beaming factor $\delta$ (to note that $\gamma_{min,max}$ of the electron energy distribution are not important in SED calculation[45]). In our calculation, the synchrotron self absorption and the Klein–Nishina effect in SSC are considered.[47,48] Lähteenmäki et al.[49] and Valtaoja et al.[24] presented the variability time scale $t_{var}$=10.7 d. We use this to determine the size of the emission region $R\sim ct_{var}\delta/(1+z)$. Other parameters can be determined by SED modeling.[45]

From AGN to the Earth, γ-ray emissions would be absorbed by extragalactic background light (EBL) through $\gamma+\gamma\rightarrow e^{\pm}$. Because of the difficulty to directly measure the EBL, several models account for the EBL spectra and its evolution.[50-52] There was 3C 279 ($z$=0.536) detected with TeV emission, which seems to constrain the EBL approaching the low limit derived from galaxy counts.[20] However Stecker et al.[53] argued that the EBL can not be constrained within current TeV data. Therefore, we use the baseline model in Stecker et al.[50] to calculate the EBL absorption. Alternatives to the model are left to be discussed.

Firstly, we model the quasi-simultaneous SED. The results are presented in Fig. 1 (the middle lines, with $\gamma_p=\gamma_0\equiv 5464$). The dotted line takes into account the EBL absorption, while the solid line does not. It can be seen (the radio emissions come from more extended region, thus it can not be fitted with our model [54,55]) that the predicted TeV flux does not reach the sensitivity of H.E.S.S., which is one of the most sensitive telescopes working at TeV band.[56] Observations of OJ 287 showed that there is one knot born each year, and the event corresponds to a radio and optical flare.[23] Thus, the long term variability (12-y period) may correspond to different emissions of various knots (also see discussion in Ref.[37]). Therefore we take the homogeneous one zone model with various parameters to describe different SEDs in high and/or low states. With approximation of the homochromous emission,[45] the synchrotron peak frequency and luminosity follow (Thomson regime),

$$\nu_s^p = \frac{2e}{3\pi mc}B\gamma_p^2\delta \propto B\gamma_p^2\delta, \qquad (2)$$

$$\nu_s^p L(\nu_s^p) = \frac{\sigma_T c}{9}R^3 B^2 \gamma_p^{3-p_1} N_0 \delta \propto R^3 B^2 \gamma_p^{1.2} N_0 \delta. \qquad (3)$$

In the last operation of Eq. (3), we use $p_1$=1.8 (see Fig. 1). The long term variability of optical emissions shows the behavior of blue-when-brighter.[31-33] This indicates that the synchrotron emission peaks at higher frequency when the source is more luminous. The x-ray study also presented evidences to this picture.[34,35,37] The similar properties are also shared by many blazars.[57] For simplicity, we assume that there are the various values for one of the jet parameters dominating this long term variable behavior. From Eqs. (2) and (3), it can be seen that this parameter can be neither the radius $R$ nor the energy density $N_0$. If it is the $B$ or $\delta$ which causes the variability, it is expected that the peak luminosity is more variable than the peak frequency (see Eqs. (2) and (3)). However, it seems that the peak frequency is more variable than the peak luminosity (see Fig. 1). Thus, it would be the different peak electron energies ($\gamma_p$) which cause the high and low states. The explanation of variability like this has also been used in other blazars (e.g., Mkn 501[58]). Therefore, we extrapolate the model to fit the variable SEDs of OJ 287 with keeping all parameters constant but $\gamma_p$.

We take $\gamma_p=\gamma_0/3$ to model the low state emissions, which are shown in Fig. 1 (the lowest black lines). It can be seen that the fittings on the low state spectra from submillimeter to x-ray bands are good. This offers an extra proof on the validity of this extrapolation modeling method. Finally, we take $\gamma_p=2\gamma_0$ to model the high state emission, which are also shown in Fig. 1 (the uppermost black lines). It can be seen that the spectra from submillimeter to x-ray can also be fitted. Taking into account the absorption by EBL, the predicted TeV flux is slightly higher than the sensitivity of H.E.S.S (see the dotted line, with a very steep photon index $\Gamma\approx4.3$ at 0.2 TeV). If we take the low-IR model for EBL absorption,[51] the calculated observational flux at TeV band would be larger than that given by the baseline model. From this aspect, the possible detection of TeV emission of OJ 287 has strong constraint on the models of EBL.

From the above analysis, we know that the outbursts at optical and x-ray bands could be indicators to the possible detection of TeV emission of OJ 287. Modeling SEDs at various epochs will constrain the relation between jet parameters at different epochs: e.g., whether $\gamma_p$ is the only parameter which dominates different emissions in high and low state. The detector Fermi/LAT would give the long-term γ-ray (GeV) variability of OJ 287 in the next few years. This combining with radio, optical and x-ray data would offer an unprecedented opportunity to explore the origin of jet activity.

In summary, we have modeled the quasi-simultaneous SEDs of LBL OJ 287. We extrapolate the model to the high state of OJ 287 to predict its γ-ray emissions, and find that its TeV emission in high state is slightly higher than the sensitivity of H.E.S.S. The next optical outburst may come at 2017,[59] which would be accompanied by a TeV outburst. Future TeV observations on OJ 287 during the high state will be very helpful to test our prediction.

We thank the anonymous referees for helpful suggestions.

**Figures Caption**

Fig. 1. SEDs of OJ 287. The red bow tie is the average spectrum from the first three-month detection by *Fermi*/LAT.[39,42] The red stars are the quasi-simultaneous multi-bands spectra within these three months operation of *Fermi*/LAT.[44] The blue squares are data from previous observations.[41,42] Three solid lines are our calculated SEDs with constant parameters but the electron peak energy $\gamma_p$. The dotted lines are the same models taking into account the absorption of γ-ray by the EBL (using the baseline model in Ref.[50]). As a comparison, the dashed line is for the sensitivity of H.E.S.S.[56]

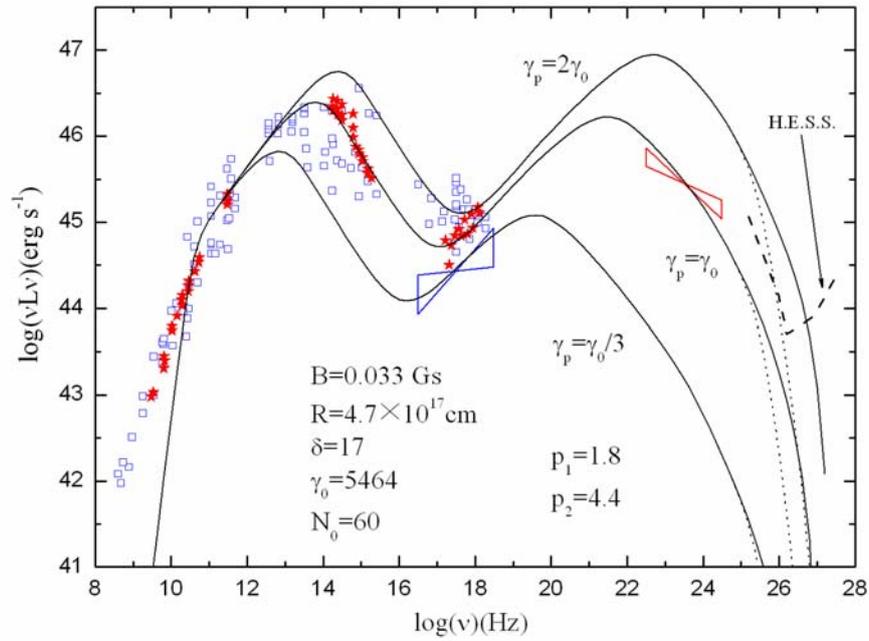

Fig. 1